\documentclass[12pt]{article}
\usepackage{pdproc,epsfig} 

\long\def\dump#1{}

\begin{document}

\newcommand{\agt}{\,\rlap{\lower3.7pt\hbox{$\mathchar\sim$}}
\raise1pt\hbox{$>$}\,}
\newcommand{\alt}{\,\rlap{\lower3.7pt\hbox{$\mathchar\sim$}}
\raise1pt\hbox{$<$}\,}

\title{AXIONS: RECENT SEARCHES AND NEW LIMITS\footnote{Based
on work done in collaboration with K.~Zioutas {\it et al.}\ (CAST
Collaboration)\cite{Andriamonje:2004hi} and with S.~Hannestad and
A.~Mirizzi\cite{Hannestad:2005df}. Contribution to Proc.\  
XI International Workshop on ``Neutrino Telescopes''
(22--25 Feb 2005, Venice, Italy).}}

\author{GEORG G.~RAFFELT}

\address{Max-Planck-Institut f\"ur Physik
(Werner-Heisenberg-Institut)\\
F\"ohringer Ring 6, 80805 M\"unchen, Germany\\
{\rm E-mail: raffelt@mppmu.mpg.de}}

\abstract{The CERN Axion Solar Telescope (CAST) experiment searches
  for solar axions by the ``helioscope'' method.  First results imply
  an upper limit on the axion-photon coupling of $g_{a\gamma} < 1.16
  \times 10^{-10}~{\rm GeV}^{-1}$ (95\% CL) for $m_a\alt 0.02~{\rm
    eV}$, in this mass range superseding the previous energy-loss
  limit from globular cluster stars.  By virtue of a variable-pressure
  helium filling of the magnetic transition region, CAST~II will
  extend the sensitivity to axion masses up to about 1~eV, for the
  first time testing realistic axion parameters in a laboratory
  experiment.  In this mass range axions would contribute a cosmic hot
  dark matter component. New structure-formation limits imply that
  $m_a<1$--2~eV. For the particular case of hadronic axions with a
  standard axion-pion coupling, the present-day cosmic axion density
  would be about $50~{\rm cm}^{-3}$ and the cosmic mass limit is
  $m_a<1.05$~eV (95\% CL).  We also comment on the axion
  interpretation of the anomalous signature observed in the PVLAS
  experiment.}

\section{Introduction}                        \label{sec:introduction}

Quantum chromodynamics is a CP-violating theory, implying that the
neutron should have a large electric dipole moment, in conflict with
the opposite experimental evidence. The most elegant solution of this
``strong CP problem'' was proposed by Peccei and Quinn (PQ) who showed
that CP conservation is dynamically restored in the presence of a new
global U(1) symmetry that is spontaneously broken at some large energy
scale\cite{Peccei:1977hh,Peccei:1977ur}.
Weinberg\cite{Weinberg:1977ma} and Wilczek\cite{Wilczek:1977pj}
realized that an inevitable consequence of the PQ mechanism is the
existence of a new pseudoscalar boson, the axion, which is the
Nambu-Goldstone boson of the PQ symmetry. This symmetry is explicitly
broken at low energies by instanton effects so that the axion acquires
a small mass. Unless there are non-QCD contributions, perhaps from
Planck-scale physics\cite{Kamionkowski:1992mf,Barr:1992qq}, the mass
is
\begin{equation}\label{eq:axmass}
m_a=\frac{z^{1/2}}{1+z}\,\frac{f_\pi m_\pi}{f_a}
=\frac{6.0~{\rm eV}}{f_a/10^6~{\rm GeV}}\,,
\end{equation}
where the energy scale $f_a$ is the axion decay constant or PQ scale
that governs all axion properties, $f_\pi=93$~MeV is the pion decay
constant, and $z=m_u/m_d$ is the mass ratio of the up and down quarks.
We follow the previous axion literature and assume a
value\cite{Gasser:1982ap,Leutwyler:1996qg} $z=0.56$, but note that
it could vary in the range\cite{Eidelman:2004wy} 0.3--0.7.

\clearpage

The PQ scale is constrained by various experiments and astrophysical
arguments that involve processes where axions interact with photons,
electrons, and hadrons\cite{Eidelman:2004wy,Raffelt:1999tx}.  The
interaction strength with these particles scales as $f_a^{-1}$, but
also includes significant uncertainties from model-dependent numerical
factors. If axions indeed exist, experimental and astrophysical limits
suggest\cite{Eidelman:2004wy,Raffelt:1999tx}
$f_a\agt0.6\times10^{9}$~GeV and $m_a\alt0.01$~eV.  The most
restrictive of these limits depends on the axion-nucleon interaction
that is constrained in two different ways by the observed neutrino
signal of supernova (SN) 1987A\cite{Eidelman:2004wy,Raffelt:1999tx}.
The axionic energy loss caused by processes such as $NN\to NNa$
excludes a window of the axion-nucleon interaction strength where the
axionic contribution to the loss or transfer of energy would have been
comparable to or larger than that of neutrinos. For a sufficiently
large interaction strength, axions no longer compete with neutrinos
for the overall energy transfer in the SN core, but then would cause
too many events in the water Cherenkov detectors that observed the
neutrino signal.  However, there is an intermediate range of
couplings, corresponding to $f_a$ around $10^6$~GeV, i.e.\ to an axion
mass of a few~eV, where neither argument is conclusive.  In this
``hadronic axion window,'' these elusive particles could still
exist\cite{Chang:1993gm,Moroi:1998qs} even if the SN~1987A limits are
taken at face value. But of course, the SN~1987A limits rely on the
model-dependent axion-nucleon coupling, they involve large statistical
and systematic uncertainties, and perhaps unrecognized loop-holes.
Therefore, it is prudent to consider other experimental or
astrophysical methods to corner axions in this range of parameters.

In Sec.~2 we will report on recent results and future plans of the
CAST experiment at CERN that searches for axions or axion-like
particles emitted by the Sun.  In Sec.~3 we will discuss the role of
axions as dark matter and in particular new limits on hot dark matter
axions.  In Sec.~4 we will briefly comment on the axion interpretation
of the anomalous signature found by the PVLAS experiment. In Sec.~5 we
will summarize and conclude.

\section{First results and future plans of the CAST experiment}

\begin{figure}[ht]
\begin{center}
\epsfig{file=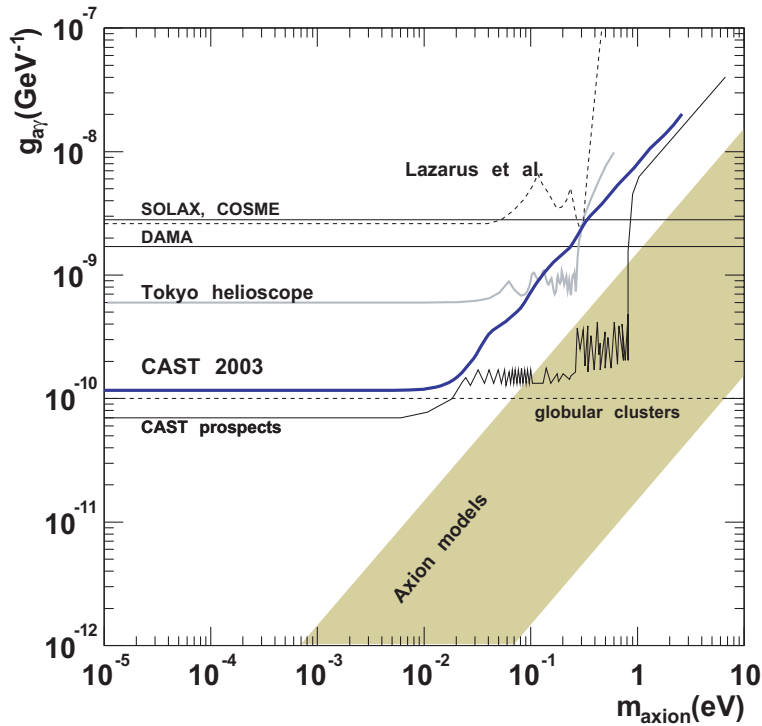,width=10cm}
\end{center}
\caption{Exclusion limit (95\% CL) from the CAST 2003 data compared
 with other constraints discussed in the text. The shaded band
 represents typical theoretical models. Also shown is the future CAST
 sensitivity as foreseen in the CAST proposal.  (Figure from the CAST
 publication\protect\cite{Andriamonje:2004hi}.)}
\label{fig:exclusion}
\end{figure}

The properties of axions are closely related to those of neutral
pions. In particular, one generic property is a two-photon interaction
of the form
\begin{equation}
{\cal L}_{a\gamma}=
-\frac{1}{4}g_{a\gamma} F_{\mu\nu}\tilde F^{\mu\nu}a
=g_{a\gamma}\,{\bf E}\cdot{\bf B}\,a\,,
\end{equation}
where $F$ is the electromagnetic field-strength tensor, $\tilde F$ its
dual, and ${\bf E}$ and ${\bf B}$ the electric and magnetic field,
respectively. The coupling constant is
\begin{equation}
g_{a\gamma}=-\frac{\alpha}{2\pi f_a}
\left(\frac{E}{N}-\frac{2}{3}\,\frac{4+z}{1+z}\right)\,,
\end{equation}
where $E$ and $N$ are the electromagnetic and color anomaly,
respectively, of the axial current associated with the axion.  In
grand unified models such as the DFSZ
model\cite{Zhitnitsky:1980tq,Dine:1981rt} one has $E/N=8/3$, but in
general the value of $E/N$ is not known so that for fixed $f_a$ a
broad range of $g_{a\gamma}$ values is possible\cite{Cheng:1995fd}.
In Fig.~\ref{fig:exclusion} we show $g_{a\gamma}$ as a function of
$m_a$ as a broad, shaded band that indicates a typical range for
$g_{a\gamma}$ although, in principle, $g_{a\gamma}$ can take on almost
any value for a given $m_a$. Still, the shaded band or ``axion line''
is the best-motivated region to search for axions.

\begin{figure}[ht]
\begin{center}
\epsfig{file=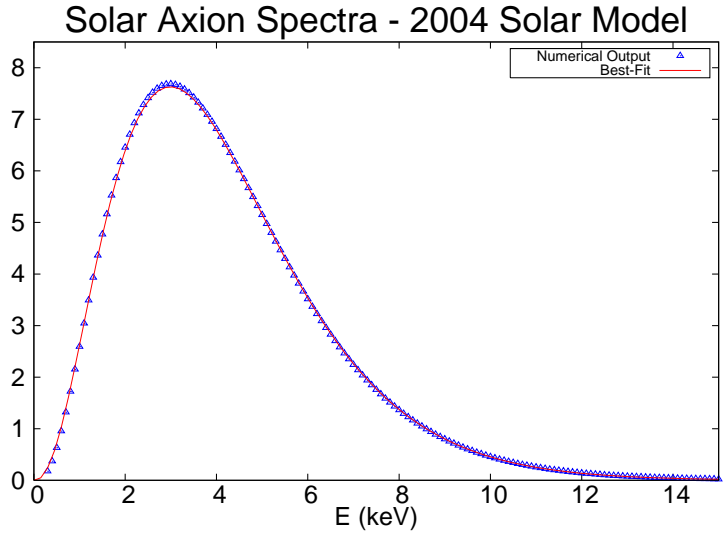,width=10cm}
\end{center}
\caption{Axion flux from a modern solar
model\protect\cite{Bahcall:2004fg} (triangles) compared with the
analytic fit Eq.~(\ref{eq:bestfit}).}
\label{fig:axflux}
\end{figure}

Particles with a two-photon interaction, and this includes gravitons
besides the hypothetical axions, can transform into photons in
external electric or magnetic fields, an effect first discussed by
Primakoff in the early days of pion physics\cite{Primakoff}.
Therefore, stars could produce these particles by transforming thermal
photons in the fluctuating electromagnetic fields of the stellar
plasma\cite{Dicus:fp}.  Calculating the expected solar axion flux is a
straightforward exercise where the only difficulty is the proper
inclusion of screening effects\cite{Raffelt:1985nk,Altherr:1993zd}. A
calculation of the axion flux expected at Earth, based on a recent
solar model\cite{Bahcall:2004fg}, is shown in Fig.~\ref{fig:axflux}. A
simple analytic fit formula is (Fig.~\ref{fig:axflux})
\begin{equation}\label{eq:bestfit}
\frac{d\Phi_a}{dE}=g_{10}^2\,\,6.020\times
10^{10}~{\rm cm}^{-2}~{\rm s}^{-1}~{\rm keV}^{-1}
\,\left(\frac{E}{\rm keV}\right)^{2.481}
\exp\left(-\frac{E}{1.205~{\rm keV}}\right)\,,
\end{equation}
where $g_{10}=g_{a\gamma}/(10^{-10}~{\rm GeV}^{-1})$.  The flux
uncertainty due to solar-model uncertainties is very small, perhaps a
few percent.

In laboratory or astrophysical $B$-fields, particles with a two-photon
vertex mix with photons so that these particles and photons
``oscillate'' much in the same way as neutrino flavor
oscillations\cite{Sikivie:ip,Raffelt:1987im}. This phenomenon can be
searched for in the laboratory\cite{Eidelman:2004wy}, can affect the
propagation of cosmic
$\gamma$-rays\cite{Gorbunov:2001gc,Csaki:2003ef}, and can modify the
apparent brightness of distant astronomical sources, in particular of
SNe Ia that are used as standard candles to measure the Hubble
diagram\cite{Csaki:2001yk,Bassett:2003zw,Ostman:2004eh}.

A particularly intriguing application of magnetically induced
axion-photon conversions is to search for solar axions by an ``axion
helioscope'' as proposed by Sikivie\cite{Sikivie:ip}.  One would look
at the Sun through a ``magnetic telescope'' and place an x-ray
detector at the far end. The axion-photon conversion probability in a
$B$-field is\cite{Sikivie:ip}
\begin{equation}
P_{a\to\gamma}=\left(\frac{g_{a\gamma}B}{q}
\right)^2\sin^2\left(\frac{qL}{2}\right)\,,
\end{equation}
where $L$ is the path length and $q$ is the axion-photon momentum
difference that in vacuum is $q=m_a^2/2E$.  For $q L\alt 1$ the
axion-photon oscillation length exceeds $L$. In the limit $q L\ll 1$
we have $P_{a\to\gamma}=(g_{a\gamma}B L/2)^2$, implying an x-ray flux
of
\begin{equation}
\Phi_\gamma=0.51~{\rm cm^{-2}~day^{-1}}\,g_{10}^4\,
\left(\frac{L}{9.26~\rm m}\right)^2 \left(\frac{B}{9.0~\rm 
T}\right)^2\,.
\end{equation}
For $qL\agt 1$ this rate is reduced due to the axion-photon momentum
mismatch. A low-$Z$ gas would provide a refractive photon mass
$m_\gamma$ so that $q=|m_\gamma^2-m_a^2|/2E$. For $m_a\approx
m_\gamma$ the maximum rate can thus be
restored\cite{vanBibber:1988ge}.

A first experiment of this sort was implemented at
Brookhaven\cite{Lazarus:1992ry}.  Later, a fully telescopic axion
helioscope with $L= 2.3~{\rm m}$ and $B=3.9~{\rm T}$ was built in
Tokyo\cite{Moriyama:1998kd}. The absence of events above background
implied $g_{10}<6.0$ at 95\%~CL for $m_a\alt 0.03~{\rm eV}$.  Later
the $m_a$ range was extended to almost 0.3~eV with a filling of helium
gas at variable pressure, reaching limits in the
range\cite{Inoue:2002qy} $g_{10}<6.8$--10.9
(Fig.~\ref{fig:exclusion}). However, this experiment did not reach the
``axion line.''

Solar axions could also transform efficiently in electric crystal
fields when the Bragg condition is satisfied\cite{Paschos:1993yf}.
However, the limits obtained by the dark matter experiments SOLAX in
Sierra Grande\cite{Avignone:1997th}, COSME in the Canfranc underground
laboratory\cite{Morales:2001we}, and DAMA at Gran
Sasso\cite{Bernabei:ny} are less restrictive than the Tokyo limit and
they also require an unacceptably high solar axion
luminosity\cite{Schlattl:1998fz}.

In order to search for solar axions an axion helioscope was built at
CERN by refurbishing a de-commissioned LHC test
magnet\cite{Zioutas:1998cc} which produces a magnetic field of
$B=9.0~\rm T$ in the interior of two parallel pipes of length
$L=9.26~\rm m$ and a cross-sectional area $A=2\times 14.5~{\rm cm}^2$.
The aperture of each of the bores fully covers the potentially
axion-emitting solar core of about one tenths of the solar radius. The
magnet is mounted on a platform with $\pm 8 ^\circ$ vertical movement,
allowing for observation of the Sun for 1.5 h at both sunrise and
sunset. The horizontal range of $\pm 40 ^\circ$ encompasses nearly the
full azimuthal movement of the Sun throughout the year.  A full
cryogenic station\cite{Barth:2004cx} is used to cool the
superconducting magnet down to 1.8~K.  The tracking system has been
precisely calibrated by means of geometric survey measurements in
order to orient the magnet to any given celestial coordinates. The
overall CAST pointing precision is better than\cite{filming}
0.01$^\circ$.

Three different detectors have searched for excess x-rays from axion
conversion in the magnet when it was pointing to the Sun. Covering
both bores of one of the magnet's ends, a conventional Time Projection
Chamber is looking for x-rays from ``sunset'' axions.  At the other
end, facing ``sunrise'' axions, a smaller gaseous chamber with novel
MICROMEGAS (micromesh gaseous structure)\cite{Giomataris:1995fq}
readout is placed behind one of the magnet bores, while in the other
one a focusing x-ray mirror telescope is working with a Charge Coupled
Device (CCD) as the focal plane detector. Both the CCD and the x-ray
telescope are prototypes developed for x-ray astronomy\cite{abrixas}.

CAST operated for about 6~months from May to November in 2003.  An
important feature of the CAST data treatment is that the detector
backgrounds are measured with about 10 times longer exposure during
the non-alignment periods.  These data are used to estimate and
subtract the true experimental background during solar tracking.  The
non-observation of a signal above background in all three detectors
leads to the exclusion range shown in Fig.~\ref{fig:exclusion}, taken
from the recent CAST publication\cite{Andriamonje:2004hi}.  In the
mass range $m_a\alt0.02$~eV the new limit (95\% CL) is
\begin{equation}
g_{a\gamma}<1.16\times 10^{-10}~{\rm GeV}^{-1}\,.
\end{equation}
In this mass range the limit is mass-independent because the
axion-photon oscillation length far exceeds the length of the magnet.
The CAST limit is far more restrictive than any previous laboratory
limit. It is comparable to the constraint obtained by the requirement
that horizontal-branch stars in globular clusters do not lose too much
energy in the form of axions\cite{Raffelt:1999tx} (``globular
cluster'' line in Fig.~\ref{fig:exclusion}).  This astrophysical
argument has not been developed to the point where it could be
associated with a precise statistical and systematic confidence level.
Therefore, in the relevant mass range the new CAST result supersedes
the globular-cluster limit even though the two constraints are
nominally comparable.

The data taken in 2004 have not yet been fully analyzed. However, the
stable operation of the experiment allowed the CAST collaboration to
take enough high-quality data to anticipate that the final sensitivity
(Fig.~\ref{fig:exclusion}) will be close to the one projected in the
CAST proposal.

The existing CAST limits and foreseen sensitivity with the 2004 data
do not yet touch the ``axion line'' in Fig.~\ref{fig:exclusion}. In
Phase~II the experiment will be modified to allow for a
variable-pressure helium filling of the magnet's bores to provide the
photons with an effective mass. The vapor pressure of He$^4$ at
1.8~K, the magnet's operating temperature, is such that a maximum
axion mass of 0.26~eV can be reached.  The pressure settings will be
incremented in steps to achieve overlapping resonant sensitivity
curves for different mass values. Using He$^3$ that has a higher
vapor pressure, one can reach a higher axion mass of up to 0.8~eV.
Reaching yet higher masses will require an isolating gas cell in the
bore where He$^3$ at 5.4~K would allow one to reach a mass of 1.4~eV.
The sensitivity forecast of CAST~II is also indicated in
Fig.~\ref{fig:exclusion}. For the first time, a laboratory experiment
will be able to probe the theoretically motivated range of axion
parameters.

\section{Axions as dark matter}

Intriguingly, axions with a mass in the eV range as probed by CAST~II
would contribute a cosmic hot dark matter component much like
neutrinos in this mass range. Of course, the exact cosmic number
density of axions depends on their primordial freeze-out epoch.
However, assuming that they freeze out after the QCD epoch, their
number density relative to a single neutrino degree of freedom will
not be diluted very much.  Therefore, the usual cosmological
structure-formation limits to neutrino masses\cite{Hannestad:2004nb}
can be generalized to other particles that where once in thermal
equilibrium\cite{Hannestad:2003ye} and particularly to
axions\cite{Hannestad:2005df}.

For any particle that thermally decouples in the early universe when
it is still relativistic, the present-day number density depends only
on the number $g_{*S}$ of effective cosmic thermal degrees of freedom
contributing to the entropy density at the decoupling epoch.  As the
universe cools, the entropy of the interacting species eventually ends
up in the cosmic microwave photons and in neutrinos, heating these
species relative to the one that is already frozen
out\cite{Kolb:1990vq}.  For a boson with a single spin degree of
freedom, the present-day number density is
\begin{equation}
n_a=\frac{g_{*S}({\rm today})}{g_{*S}({\rm decoupling})}\times
\,\frac{n_\gamma}{2}\,,
\end{equation}
where $n_\gamma=411~{\rm cm}^{-3}$ is the present-day density of
cosmic microwave photons and today\cite{Kolb:1990vq} $g_{*S}=3.91$.
For cosmic epochs before neutrino decoupling, the effective number of
thermal degrees of freedom $g_*$ characterizing the energy density and
$g_{*S}$ for the entropy density are almost identical so that
henceforth we will not distinguish between the two quantities.

\begin{figure}[ht]
\begin{center}
\epsfig{file=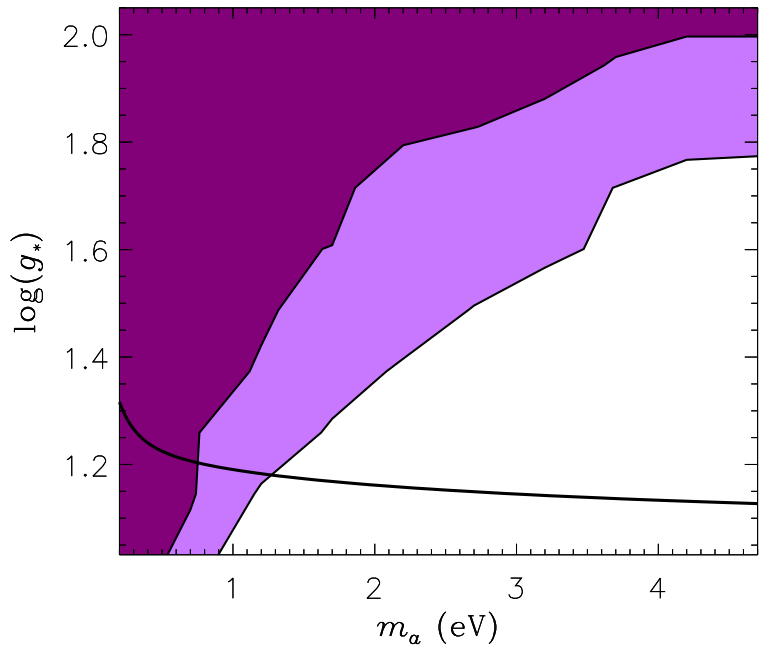,width=10cm}
\end{center}
\caption{Likelihood contours for the cosmologically allowed axion
  mass $m_a$ and $g_*$, the number of effective cosmic thermal degrees
  of freedom at the epoch of axion decoupling. Everything to the right
  of the dark shaded region is excluded at the 68\% CL, and everything
  to the right of the light shaded region is excluded at the 95\% CL.
  The relation between $g_*$ and $m_a$ for hadronic axions is shown as
  a thick solid line.  (Figure from Hannestad, Mirizzi and
  Raffelt\protect\cite{Hannestad:2005df}.)}
\label{fig:cosmiclimits}
\end{figure}

Once the decoupling epoch for the new particles has been established,
their number density and velocity dispersion are known so that one can
determine if the corresponding hot dark matter fraction is compatible
with the usual large-scale structure and cosmic microwave background
radiation data.  In Fig.~\ref{fig:cosmiclimits} we show the allowed
range for $m_a$ and $g_*$ from the analysis of Hannestad, Mirizzi and
Raffelt\cite{Hannestad:2005df}.

At the time of neutrino decoupling at a temperature of about 1~MeV,
only photons, electrons and neutrinos contribute to the cosmic
radiation density so that $g_*=10.75$. At higher temperatures, muons,
pions and other mesons begin to contribute so that $g_*$ increases to
about 23 at $T_{\rm QCD}\approx150$~MeV when the color deconfinement
transition occurs, causing a fast increase of $g_*$ to about 100 for
$T>T_{\rm QCD}$. Therefore, if axions thermally decouple after the QCD
epoch, their mass can not exceed 1--2~eV to avoid too much
free-streaming erasure of small scale structure in the universe.

If we consider new axion-like particles that have no other
interactions at low energies besides the two-photon vertex, we can
estimate their freeze-out temperature by the simple criterion that
their interaction rate $\Gamma$ should exceed the cosmic expansion
rate $H$.  On dimensional grounds we have $H\sim T^2/m_{\rm Pl}$ with
$m_{\rm Pl}=1.2\times10^{19}~{\rm GeV}$ the Planck mass and
$\Gamma\sim g_{a\gamma}^2T^3$.  Therefore, $\Gamma\sim H$ implies
$T_{\rm decoupling}\sim g_{a\gamma}^{-2} m_{\rm Pl}^{-1}$.  With
$g_{a\gamma}\alt 10^{-10}~{\rm GeV}^{-1}$ we have $T_{\rm
  decoupling}\agt 10~{\rm GeV}\gg T_{\rm QCD}$ and thus $g_*>100$--200
at decoupling.  Therefore, we have no substantial mass limit on such
particles.

Turning to realistic models in the context of the PQ mechanism, axions
generically interact with hadrons, even though the exact coupling
constants are model dependent. In particular, the axion-pion
interaction is of the form\cite{Chang:1993gm}
\begin{equation}\label{eq:axionpionlagrangian}
{\cal L}_{a\pi}=\frac{C_{a\pi}}{f_\pi f_a}\,
\left(\pi^0\pi^+\partial_\mu\pi^-
+\pi^0\pi^-\partial_\mu\pi^+
-2\pi^+\pi^-\partial_\mu\pi^0\right)
\partial_\mu a\,.
\end{equation}
In hadronic axion models where the ordinary quarks and leptons do not
carry PQ charges, the coupling constant is\cite{Chang:1993gm}
\begin{equation}\label{eq:axionpioncoupling}
C_{a\pi}=\frac{1-z}{3\,(1+z)}\,.
\end{equation}
In non-hadronic models an additional term enters that could reduce
$C_{a\pi}$.

Assuming the hadronic axion-pion interaction strength, the decoupling
temperature and the corresponding $g_*$ was calculated on the basis of
the dominant $\pi\pi\leftrightarrow\pi a$
processes\cite{Hannestad:2005df,Chang:1993gm}.  Assuming in addition
the standard $m_a$-$f_a$ relation of Eq.~(\ref{eq:axmass}), hadronic
axions live on the thick solid line shown in
Fig.~\ref{fig:cosmiclimits}. In this case the axion parameter space
collapses to one dimension, i.e.\ the cosmological maximum likelihood
analysis involves only a single independent axion parameter. In this
case one finds\cite{Hannestad:2005df}
\begin{eqnarray}
m_a&<&1.05~{\rm eV}\,,\nonumber\\
f_a&>&5.7\times10^6~{\rm GeV}
\end{eqnarray}
at 95\%~{CL}.  The present-day cosmic axion density would be about
$50~{\rm cm}^{-3}$.  For comparison we note that the same method
provides\cite{Hannestad:2004bu} $\sum m_\nu < 0.65~{\rm eV}$ at
95\%~{\rm CL}. It is understood that these constraints involve a
number of well-recognized systematic uncertainties inherent in the
underlying cosmological assumptions.

Axions are an often-cited cold dark matter candidate if they interact
so weakly (if $f_a$ is so large) that they never achieve thermal
equilibrium\cite{Preskill:1982cy,Abbott:1982af,Dine:1982ah,%
  Davis:1986xc}. In that case coherent oscillations of the axion field
are excited around the QCD transition when the PQ symmetry is
explicitly broken by instanton effects. The mass range where axions
would contribute the cosmic cold dark matter is fairly uncertain but
probably is somewhere in the range\cite{Eidelman:2004wy} $1~\mu{\rm
  eV}<m_a<1~{\rm meV}$.  Galactic dark matter axions in the lower
range of these plausible masses are searched for by the well-known
microwave cavity experiments\cite{Bradley:2003kg}. The main point for
our present discussion is that axions, depending on their mass, can
contribute to the cosmic cold dark matter or the hot dark matter.  It
is important to note that very low-mass axions would be cold dark
matter, whereas eV-mass axions would be a hot dark matter component.
This situation is opposite for neutrino-like particles (WIMPs) that
contribute to the hot dark matter for eV-range masses whereas they are
cold dark matter typically with masses beyond several tens of GeV. The
key difference to axions is that axions can be a cold dark matter
candidate only if they never reached thermal equilibrium and thus
would survive as a non-thermal relic.

If CAST~II were to detect axions in the sub-eV mass range, these
particles would contribute a new hot dark matter component in addition
to the one provided by ordinary neutrinos. At the same time axions
would then be excluded as a candidate for cold dark matter.

\section{Axion interpretation of the PVLAS signature?}

Another possibility to search for particles with a two-photon vertex
is their effect on the propagation of a laser beam in the presence of
a transverse magnetic field.  The photon state polarized
perpendicularly to ${\bf B}$ remains unaffected whereas the parallel
one undergoes partial photon-axion oscillations.  As a result, a
potentially measurable effect on the overall polarization of the laser
beam occurs\cite{Maiani:1986md,Raffelt:1987im}.  The PVLAS
experiment\cite{Gastaldi} is an ongoing effort to search for this
effect. Intriguingly, a strong effect has been observed that at
present can not be ascribed to any known instrumental factors and thus
could be indicative of new physics.

One tentative interpretation is that PVLAS may actually be observing
axion-like particles. The required axion-photon coupling strength is
around $3\times10^{-6}~{\rm GeV}^{-1}$ and thus much larger than what
is allowed by either the CAST limits, the globular-cluster constraint,
or simply the observed properties of our Sun\cite{Schlattl:1998fz}.
In that sense CAST does not directly rule out the PVLAS effect because
CAST relies on the standard Sun as a source where axion emission is
only a small effect. We note that the solar axion luminosity due to
the Primakoff effect is $L_a=g_{10}^2\,1.7\times10^{-3}\,L_\odot$ so
that the PVLAS-suggested value for $g_{a\gamma}$ implies an axion
luminosity about a million times larger than the solar photon
luminosity.  Such a large effect can not be accommodated in a
self-consistent solar model and in any case, the Sun would consume its
nuclear fuel a million times faster and could only exist for a few
thousand years.

One may think that this argument can be circumvented by assuming that
the PVLAS-particles interact so strongly in some reaction channel that
they can not freely escape from the Sun. This assumption, however,
does not provide a viable loop hole. Even if the particles can not
freely escape, they will still contribute to the transfer of energy
within the Sun and thus strongly affect the Sun's structure and its
speed of energy loss. A low-mass particle would be harmless only if
its interaction cross section is either so small that it does not
carry away too much energy or else more strongly interacting than
electromagnetic strength to avoid excessive energy transfer in the
Sun\cite{Raffelt:1988rx}.  In this latter case presumably it would not
have escaped experimental detection.

Therefore, the axion interpretation of the PVLAS signature does not
appear viable unless one can come forth with a self-consistent
mechanism that avoids excessive energy loss or energy transfer in the
Sun and other stars. A first attempt to construct a particle-physics
model for the PVLAS signature and yet avoids the solar energy transfer
argument has just been reported\cite{Masso:2005ym}. If it is viable
remains to be discussed.

\eject

\section{Summary}

First results from the CAST experiment, based on the 2003 data,
provide a new limit on the two-photon vertex of axions or other
similar hypothetical particles.  In the mass range $m_a\alt0.02$~eV
the new limit supersedes the previous globular-cluster constraint.
The 2004 data will provide significantly improved sensitivity. In
phase~II, CAST will use a variable-pressure buffer gas in the magnetic
transition region to increase its sensitivity to axion masses up to
around 1~eV. If axions were discovered in this mass range they would
contribute a cosmic hot dark matter component in addition to ordinary
neutrinos.  Conversely, structure-formation limits imply that hot dark
matter axions should obey an upper mass limit of 1--2~eV, depending on
the exact axion-pion interaction strength. The axion interpretation of
the PVLAS experiment does not appear viable in view of the properties
of our Sun. Therefore, the search for axions continues, both by the
CAST experiment for solar axions and by the cavity searches for
galactic cold dark matter axions.

\section{Acknowledgments}

I thank my collaborators in the works described here, i.e.\ K.~Zioutas
{\it et al.}  (CAST Collaboration)\cite{Andriamonje:2004hi} and
S.~Hannestad and A.~Mirizzi\cite{Hannestad:2005df}.  Partial support
by the Deutsche Forschungsgemeinschaft under grant No.~SFB 375 and by
the European Union under the Ilias project, contract
No.~RII3-CT-2004-506222, is acknowledged.


\end{document}